\documentclass[aps,prb,reprint,groupedaddress,twocolumn]{revtex4}
\usepackage{amsmath,amssymb,braket,mathtools,array,MnSymbol,bm}
\usepackage{graphicx,color}
\usepackage{braket}

\usepackage[thinlines]{easytable}

\usepackage{hyperref}
\hypersetup{
	colorlinks   = true, 
	urlcolor     = blue, 
	linkcolor    = blue, 
	citecolor   = blue 
}

\graphicspath{{images/}}

\newcommand{\be}{\begin{equation}}
\newcommand{\ee}{\end{equation}}
\newcommand{\bea}{\begin{eqnarray}}
\newcommand{\eea}{\end{eqnarray}}
\newcommand\ba{\begin{align}}
\newcommand\ea{\end{align}}

\newcommand{\abs}[1]{\lvert#1\rvert}

\renewcommand{\vec}[1]{{\boldsymbol #1}}

\renewcommand{\Im}{{\rm \, Im\,}}
\newcommand{\tr}{{\rm tr\,}}
\newcommand{\Tr}{{\rm Tr\,}}

\newcommand{\A}{{\mathcal A}}

\newcommand{\lb}{\left[}
\newcommand{\rb}{\right]}

\begin{document}

\title{Ultrasonic attenuation in a pseudogapped superconductor}

\author{A. V. Shtyk $^{1,2}$ and  M. V. Feigel'man$^{3,4}$}

\affiliation{$^{1}$ Physics Department, Harvard University, USA}

\affiliation{$^{2}$ Moscow Institute of Physics and Technology, Dolgoprudny,
Moscow region, Russia}

\affiliation{$^{3}$ L. D. Landau Institute for Theoretical Physics, Chernogolovka, 142432, Moscow region, Russia}

\affiliation{$^{4}$ Laboratory for Condensed Matter Physics, Higher School of Economics, Moscow, Russia}

\begin{abstract}
We develop a theory of ultrasound decay rate $\alpha$  in a model of a superconductor with a large pseudogap $\Delta_P$, 
and show that at low temperatures ($T \ll T_c$) the  magnitude of the decay rate $\alpha$ is 
controlled by the ratio of $T/\Delta$, where $\Delta \ll \Delta_P$ is a superconducting collective gap. 
Thus we propose  new method to measure the collective gap $\Delta$ in a situation when
 strong pseudogap is present.

\end{abstract}

\date{\today}
\maketitle

\section{Introduction}
Strongly disordered superconductors near quantum phase transition QPT) into an insulator state became again an object of 
great interest during last decade, both on experimental side~\cite{e1,e2,e3,e4,NP2011,SacepeTiN,Sacepe2009,N1,Teun1} 
and among theorists~\cite{t1,t2,t3,t4} (references given above are certainly incomplete, due to a large number of papers
in the field). 
 On experimental side, revival of 
the interest to this subject comes about due to new methods which became available. In particular, 
low-temperature scanning tunneling spectroscopy  makes it possible to study properties of 
superconducting state locally with a nanometer-scale resolution, which allowed to demonstrate~\cite{NP2011,SacepeTiN} 
an existence of a strong density-of-states (DoS) suppression  at temperatures much above the superconducting transition
 $T_c$. In particular, amorphous InO$_x$ films demonstrate virtually zero DoS at and around Fermi level at $T \leq 1.5 T_c$.
Such a phenomenon was called \textit{pseudogap}, in analogy to the somewhat similar phenomenon known for under-doped
 high-$T_c$ oxide superconductors. Experimentally, clear distinction between single-particle gap (pseudo-gap) $\Delta_P$
 and collective superconducting gap $\Delta$ was made by means of Andreev contact spectroscopy~\cite{A1} of 
the same InO$_x$ films.

 Theoretically, it became possible to understand~\cite{t1}  the origin of a pseudogap as a result of an effective 
(phonon-mediated) electron-electron attraction acting between Anderson-localized electrons. A detailed  semi-quantitative
theory of  superconductivity, starting from BCS-like model with localized single-electron states (near 3D Anderson 
localization transition) was developed in Ref.~\cite{t2},  elaborating an approach proposed originally in
\cite{ML} and developed numerically in~\cite{Ghosal}.
 Qualitatively similar results were later obtained~\cite{tb1,tb2,tb3}
 by means of Renormalization Group methods developed for 2D systems. Application of  ideas developed in ~\cite{t2} was 
found to be useful~\cite{FI2015} to the understanding of unusual scaling of superconducting density versus superconducting gap,
 $\rho_s \propto \Delta^2$, as reported in Ref.\cite{microwaves2015}.  A microwave absorption technique used 
in \cite{microwaves2015} may present an alternative to Andreev spectroscopy~\cite{A1,A2} for measurement of the genuine
superconducting gap.  

The present paper is devoted to development of a theory of still another phenomenon that is directly related to the
collective superconducting gap in disordered materials  at very low temperatures near SIT. 
Namely, we consider attenuation rate $\alpha(\omega,T)$ of ultrasound wave propagating in a superconducting material
with a well developed pseudogap $\Delta_P \gg \Delta$, at the temperature range $T \ll \Delta/k_B$ and at 
relatively low frequencies $\hbar \omega \ll k_B T $ \, (below the Boltzmann's constant $k_B$ is set to unity).
In the paper~\cite{SKF} we have shown that ultrasound attenuation rate is intrinsically related to the 
electron-phonon inelastic energy exchange, which becomes very inefficient at sub-Kelvin temperatures, leading
to thermal instabilities and electron overheating~\cite{Kravtsov2009} in a "neighbouring" \, insulating state of 
the same (or similar) materials~\cite{Sacepe2009,Ladie,new}.   We identified in Ref.~\cite{SKF} a few mechanisms
 leading  to (possibly strong) enhancement of the electron-phonon inelastic coupling due to the presence of 
slowly diffusing modes which may exist in an electronic liquid (particle density, magnetization density, etc.).

In Ref.~\cite{Att1} we have studied one more example of such a diffusion-controlled enhancement, now
 due to diffusion of thermal energy.   This mechanism is quite universally present in any disordered conductor,
 while its relative strength (with respect to the standard local Pippard mechanism~\cite{Pippard}) may vary a lot.
 In particular, we have calculated~\cite{Att1} ultrasound attenuation $\alpha(\omega,T)$
  for both  s-wave and d-wave BCS-type superconducting states (as well as in normal doped Silicon);  
in the s-wave case both Pippard's and diffusion-controlled mechanisms provide an exponentially suppressed 
low-temperature  attenuation $\alpha(\omega,T) \propto exp(-\Delta/T)$, differing in preexponential factors. 

For a pseudogapped superconductor with large single-particle gap $\Delta_P$, all \textit{local} inelastic 
electron-phonon processes are obviously suppressed $\propto exp(-\Delta_P/T)$, while ultrasound 
attenuation due to energy diffusion mode are expected to contain exponential factors  $\propto exp(-\Delta/T)$,
with smaller collective gap $\Delta$.  Thus it is natural to expect that low-temperature behavior of
$\alpha(\omega,T)$ will be controlled by this latter mechanism.  The purpose of the present paper is to 
calculate $\alpha(\omega,T)$ within the simplest model of a pseudogapped superconductor.
The rest of the paper is organized as follows: in Sec.II we formulate our basic model for a superconductor with
a large pseudogap  and discuss its basic properties, Sec. III is devoted to the derivation of the form of collective
 modes present in the superconducting state,  then in Sec. IV  we introduce simplest model of electron-phonon
 coupling relevant for the pseudogapped superconductor.  Sec. V contains derivation of our main results: 
ultrasonic attenuation rate $\alpha(\omega,T)$  due to coupling between phonons and two collective superconducting
 modes (phase and amplitude modes) is derived.  Sec. VI is devoted to a  discussion of the role of long-range Coulomb 
 interaction between electron pairs.
 Finally, Sec. VI contains our conclusions.

\section{Model Hamiltonian and effective action}

According to the analysis developed in Ref.~\cite{t2}, pseudogapped superconducting state can be realized in 
poor conductors where single-particle electron states are weakly localized (localization length $L_{loc}$ is larger than
mean distance between electrons which eventually participate in superconductivity), and effective Cooper  attraction
 is present between electrons.
For the last condition to be realized  together with sufficient disorder needed for localization,  
Coulomb repulsion between conduction electrons should be strongly suppressed. This is the reason to start from
the simplest model formulated in terms of "pseudo-spins"  $1/2$ introduced originally by P.W.Anderson in~\cite{Anderson59}:
\begin{equation}
\label{H0}
	H[\vec S_i]=-2\sum_{i=1}^{N} \xi_iS_i^z-\sum_{i>j=1}^{N}J_{ij}\lb S_i^xS_j^x+S_i^yS_j^y\rb,
\end{equation}
Here $S^+_i = S^x_i + i S^y_i$ and $S^-_i = S^x_i - i S^y_i$  are  operators which create (annihilate) a pair
 of electrons populating $i$-th  localized eigenstate of the free-electron problem, while  $S^z_i + 1/2$ 
counts the number of these pairs. Each localized eigenstate can be  characterized by location of its maxima and by  its eigenvalue
$\xi_i$. Correspondingly, $2\xi_i$ is a local  energy of a Cooper pair siting on a "site" \, $i$.
For simplicity, we assume sites $i$ to be  arranged into cubic lattice with an elementary cell of size $a$.
Hamiltonian (\ref{H0}) acts in the reduced Hilbert space, spanned by localized electron pairs with zero total spin; 
single-electron population of any localized state is excluded, due to  high extra 
energy $\Delta_P \gg \Delta$ associated with it ( parity gap, see Ref.~\cite{MatveevLarkin}).

Magnitudes of $\xi_i$  are random with a distribution function $p(\xi)$. 
 We assume  a box-shaped  $p(\xi)=(2W)^{-1}\Theta(W-\abs{\xi})$  with energies in the interval $\xi\in(-W,W)$,
 although any distribution which is flat around Fermi energy position will lead to the same physical results.
Then effective density of states (DoS) at the Fermi level is given by $\nu = 1/2Wa^3$.
 The second term of Eq.(\ref{H0}) describes hopping of Cooper pairs between the orbitals, that is equivalent to interaction
between pseudospins. In the long-wavelength limit this (Fourier-transformed) interaction can be expanded in powers of small momenta $\vec{k}$,
\begin{equation}
\label{J}
	J(\vec{k})=g\left[1-R^2k^2+O(k^4)\right],
\end{equation}
where $g\equiv J(\vec{0}) \sim J_{ij} \cdot R^3$  is the overall coupling strength and $R$ can be interpreted as a typical interaction range 
(i.e. Cooper pair hopping range). According to Ref.~\cite{t2}, we expect $R$ to be somewhat larger than $L_{loc}$.
The Hamiltonian (\ref{H0}) does not contain any remnants of  long-range Coulomb repulsion 
between electrons; this is an idealized model which we are going to start from. We will discuss the role of (weak) 
Coulomb  repulsion in the end of the paper.  The same Hamiltonian (\ref{H0}) was employed in the paper~\cite{FIM2010}
to study quantum phase transition (QPT)  between superconducting state (in spin terms, it is the state with 
nonzero average $\langle S^{x,y}_i \rangle$) and insulating state.  In the present paper we will not consider 
specific features related to this  QPT; instead, we are interested in the properties of collective excitations within
 well-developed superconducting state.  This is why we choose to work here with the model  of very large interaction 
(hopping) range $R \gg a$, and we will employ further a mean-field approximation based on this inequility.
In this sense, our approach is similar to the usual semiclassical theory of superconductivity.

On a technical side, we  choose Fedotov-Popov representation~\cite{FedotovPopov} for spin-$\frac12$ operators, 
that is useful to construct a diagrammatic approach and to study collective modes in the ordered state. 
It is shown in paper~\cite{FedotovPopov} (see also some extension of this approach in~\cite{KiselevOppermann})
that exact representation of the  partition function for interacting spin systems can be obtained by the
representation of spin operators via a  special kind of fermionic operators:
\begin{equation}
S_i^{\alpha}=(1/2)\psi^\dagger_i\sigma^\alpha\psi_i
\label{FP}
\end{equation}
where $\sigma^\alpha$ are three Pauli matrices and (anticommuting) two-component spinor operators $\psi(\tau),\psi^\dagger(\tau)$
 obey the following boundary condition in the Matsubara imaginary time: 
$\psi(\tau + \beta) = i \psi(\tau)$, $\psi^\dagger(\tau + \beta) = -i \psi^\dagger(\tau)$,where $\beta = \hbar/T$.
Following~~\cite{KiselevOppermann}, we will refer to such a modified fermions as to "semions".

Using representation (\ref{FP}), we rewrite the original Hamiltonian in the form
\begin{align}
H&=-\sum_i\psi_i^+\xi_i\sigma^z\psi_i
-\frac{1}{4}\sum_{i>j}J_{ij}(\psi^{\dagger}_i\sigma^\alpha\psi_i)(\psi_j^+\sigma^\alpha\psi_j),
\end{align}
where in the interaction term the index $\alpha=x,y$. To treat the interaction term, 
we introduce a complex order parameter field $\Delta$ via Hubbard-Stratonovich transformation
and integrate out semions. That results in the  effective imaginary time action:
\begin{equation}
\begin{split}
	\label{A0}
	\A[\Delta]=&-\Tr\lb \Delta^{*}J^{-1}\Delta\rb+
	\\
	&+\Tr\ln\lb i\varepsilon_l+\xi\sigma^z +\frac{1}{2}(\Delta \sigma^-+\Delta^* \sigma^+) \rb.
\end{split}
\end{equation}
Matsubara energies here are of semionic nature and read as $\varepsilon_l=2\pi T(l+1/4)$, while traces go over all spaces.

The equilibrium order parameter is determined by the self-consistency equation that follows from the variation of the Action,
 Eq.(\ref{A0}), over $\Delta$:
\begin{equation}
\label{SCE}
	1= \frac{g}{a^3}\int \frac{p(\xi) d\xi}{\sqrt{\xi^2+\Delta^2}}\tanh\frac{\sqrt{\xi^2+\Delta^2}}{T}
\end{equation}
This yields $\Delta_0=2We^{-1/\lambda}$ for zero-temperature order parameter and $T_c=2.27We^{-1/\lambda}$ for transition temperature. 
Here $\lambda \equiv\frac{g}{W a^3} $ is a dimensionless coupling constant.
%
Action (\ref{A0}) will be used throughout  our further analysis.

\section{Collective modes: fluctuations in the ordered state.}

In a superconducting phase, for temperatures $T<T_c$, order parameter can be parametrized as
\begin{equation}
	\Delta=(1+\eta)\Delta_0 e^{i\varphi},
\end{equation}
revealing two collective modes: a massive (amplitude or Anderson-Higgs) mode $\eta$ and a massless Goldstone boson, phase $\phi$.
Below we derive expressions for propagators of both these modes.

The action (\ref{A0}) can be expanded in fluctuations around the ground state $\Delta_0$. The quadratic Gaussian part describes dynamics of collective modes. Once expanded, the action gives the amplitude propagator
%
\begin{equation}
	L^{-1}_{\eta}(\Omega,\vec{k})=\frac{\Delta^2}{a^3}\left[-J^{-1}(\vec{k}) 
	+\Pi_{xx}(\Omega,\vec{k})\right],
\end{equation}
where $\Pi_{xx}$ is a $\sigma_x-\sigma_x$ semionic correlator,
\begin{align}
\Pi_{xx}=\sumint_{l,\xi}\tr\sigma_x G(i\varepsilon_l)\sigma_x G(i\varepsilon_{l+n}).
\label{Pix}
\end{align}
Here the sign $\sumint_{l,\xi}=T\sum_l\int(d\xi/2W)$ and $G$ is a basic building block of the diagrammatic technique, a semionic Green function that can be represented as
\begin{align}
G(i\varepsilon_l)=\frac{1}{i\varepsilon_l+\vec{E}\cdot\vec{\sigma}}=\sum_{\pm}\left(\frac{1}{2}\pm\frac{\vec{E}\cdot\vec{\sigma}}{2E}\right)\frac{1}{i\varepsilon_l\pm E}
\end{align}
with a vector $\vec{E}=(\Delta,0,\xi)$, $E=\sqrt{\Delta^2+\xi^2}$.

Calculation of the correlator gives
\begin{align}
\Pi_{xx}=\sumint_{l,\xi}&\left[\frac{\Delta^2}{E^2}\left(\frac{1}{i\varepsilon_{l+n}-E}\frac{1}{i\varepsilon_{l}-E}+(E\rightarrow-E)\right)+\right.
\\  \nonumber
&+\left.\frac{\xi^2}{E^2}\left(\frac{1}{i\varepsilon_{l+n}+E}\frac{1}{i\varepsilon_{l}-E}+(E\rightarrow-E)\right)\right],
\end{align}
where in the low-temperature $T\ll\Delta$ limit the first term gives a negligible exponentially small contribution, while the second term leads to
\begin{align}
	\Pi_{xx}=&\int\frac{p(\xi) \xi^2 d\xi}{E^2}\cdot\frac{E}{E^2+(\Omega_n/2)^2}
	\\ \nonumber
	=&g^{-1}-W^{-1}\frac{\sqrt{4\Delta^2+\Omega_n^2}}{\left|\Omega_n\right|}\text{arcsh} \left|\frac{\Omega_n}{2\Delta}\right|,
\end{align}
where we made use of the self-consistency equation (\ref{SCE}). Recalling the gradient expansion (\ref{J}) we have an imaginary time amplitude propagator
\begin{align}
\label{L_eta}
	L^{-1}_{\eta}(i\Omega_n,\vec{k})=&-\nu\left[\Delta^2b\left(\frac{i\Omega_n}{2\Delta}\right)+\frac{v^2k^2}{4}\right],
\end{align}
where we had introduced a velocity 
\begin{equation}
	v=\lambda^{-1/2}\Delta R
\end{equation}
and a function
\begin{equation}
\label{a}
	b(ix)=\sqrt{1+x^{2}} \frac{\text{arcsh} \abs{x}}{\abs{x}}.
\end{equation}

\begin{figure}[t]
	\begin{minipage}[h]{0.49\linewidth}
		\center{\includegraphics[width=0.99\linewidth]{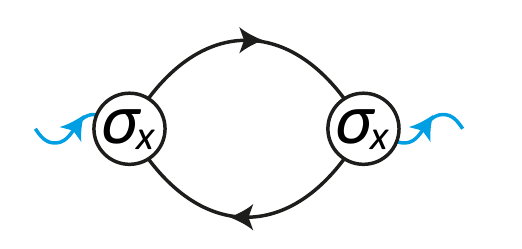}}
	\end{minipage}
	\hfill
	\begin{minipage}[h]{0.49\linewidth}
		\center{\includegraphics[width=0.99\linewidth]{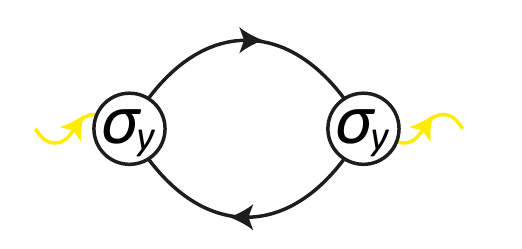}}
	\end{minipage}
	\caption{Self-energy of the collective superconducting mode (amplitude is shown on the left, with blue external lines, and phase is on the right, yellow lines). Black lines stand for semionic propagators.}
	\label{ch6:fig:Pi}
\end{figure}

The obtained propagator has branch a cut along $(-\infty,-2\Delta]\cup[2\Delta,+\infty)$ rather than a simple pole. The analytical continuation to real frequencies $i\Omega_n\rightarrow\Omega+i0$ gives for a retarded propagator
\begin{align}
	\left( L^{R}_{\eta}(\Omega,\vec{Q}) \right)^{-1} =&-\nu\left[\Delta^2b^R\left(\frac{\Omega+i0}{2\Delta}\right)+\frac{v^2k^2}{4}\right],
\end{align}
with a function
%
\begin{equation}
b^\text{R}(x)=
\begin{cases}
\sqrt{1-x^2}(\arcsin x/x) & \abs{x}<1
\\
\sqrt{1-x^{-2}}\left[\frac{i\pi}{2}+\text{arch}\, x\right] &\abs{x}>1
\end{cases}.
\end{equation}
Note that the gap edge for amplitude excitations is located at the energy $2\Delta$, unlike usual BCS superconductors where
energies of elementary fermionic excitations start from $\Delta$\, (on the other hand, minimal energy needed to 
split a Cooper pair is always equal to $2\Delta$ since two quasiparticles should be produced).

For energies just above the gap $b^\text{R}\simeq i\pi\sqrt{(x-1)/2}$, so that
\begin{equation}
\label{L_eta_2}
\begin{split}
	\left(L^{R}_{\eta}(\Omega,\vec{Q})\right)^{-1}=-\nu\Delta^2\left[i\pi\sqrt{\frac{\Omega-2\Delta}{4\Delta}}+\frac{v^2k^2}{4\Delta^2}\right],
	\\
	(\Omega> 2\Delta,\, \abs{\Omega-2\Delta}\ll\Delta)
\end{split}
\end{equation}
while for large energies $b^\text{R}\simeq i\pi/2+\ln2x$,
\begin{align}
\begin{split}
	\left(L^{R}_{\eta}(\Omega,\vec{Q})\right)^{-1}=-\nu\Delta^2\left[\frac{i\pi}{2}+\ln\frac{\Omega}{\Delta}+\frac{v^2k^2}{4\Delta^2}\right].
	\\
	(\Omega\gg 2\Delta)
\end{split}
\end{align}
Finally, for small energies we have
\begin{align}
\begin{split}
\left(L^{R}_{\eta}(\Omega,\vec{Q})\right)^{-1}=-\nu\Delta^2\left[1-\frac{\Omega^2}{12\Delta^2}+\frac{v^2k^2}{4\Delta^2}\right]
\\
(\abs{\Omega}\ll 2\Delta)
\end{split}
\end{align}
implying a zero temperature coherence length $\xi=v/2\Delta$.

Meanwhile, phase mode is gapless and relevant energies are well below the gap $\Delta$. In other words we can safely expand the propagator in the frequency $\Omega_n$. Performing a straightforward calculation and a subsequent analytical continuation we eventually have 
\begin{align}
\label{L_phi}
	\left(L^{R}_{\phi}(\Omega,\vec{k})\right)^{-1}=-\frac{\nu}{4}\left[-\Omega^2+v^2k^2\right].
\end{align}

Expressions derived in this Section for the propagators $L_\phi$ and $L_\eta$ will be used below for the calculations of 
the phonon decay rate.

\section{Electron-phonon interaction}

\begin{figure}[b]
	\begin{minipage}[h]{0.49\linewidth}
		\center{\includegraphics[width=0.99\linewidth]{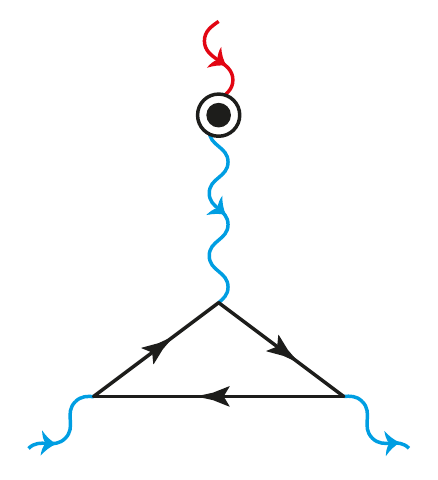}}
	\end{minipage}
	\hfill
	\begin{minipage}[h]{0.49\linewidth}
		\center{\includegraphics[width=0.99\linewidth]{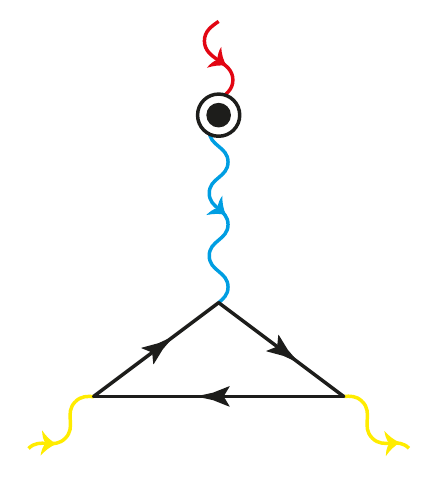}}
	\end{minipage}
	\caption{Interaction vertex of the type phonon-collective superconducting-modes. Left: conversion of a phonon into two amplitude modes. Right: conversion into two phase modes. Black lines in the triangle stand for semionic propagators.}
\end{figure}

Interaction with longitudinal phonons can be introduced via modulations of the Cooper pair hopping amplitude (i.e. pseudospin interaction constant) as
\begin{equation}
H_{\text{e-ph}}[\vec S_i]=\kappa\sum_{i,j=1}^{N}(J_{ij}\text{ div}\vec{u})\left[ S_i^xS_j^x+S_i^yS_j^y\right]
\end{equation}
with a coupling constant $\kappa$ that is normally of order of unity, $\kappa\sim1$. The choice of such a model for
 e-ph interaction in the effective Hamiltonian makes sense since the pair hopping term in the Hamiltonian originates from
 the  original phonon-mediated Cooper attraction between electrons.

In a close parallel with electron-phonon interaction in a weakly disordered superconductors, the effect of an acoustic wave  within adiabatic approximation in the limit $\omega,q\rightarrow0$ is reduced to modulations of the dimensionless interaction constant
\begin{equation}
	\lambda\rightarrow \lambda(1-\kappa\,\text{div}\vec{u}).
\end{equation}
Consecutively, changes of the coupling constant $\lambda$ are  most clearly revealed by the change in the ground state $\Delta_0$ due to the exponential sensitivity of the former,
\begin{equation}
	\Delta=2We^{-1/\lambda}\rightarrow\Delta\left(1+\frac{\kappa}{\lambda}\text{div}\vec{u}\right).
\end{equation}

Eventually this gives electron-phonon interaction
\begin{align}
\label{A_e-ph_0}
	\A_{\text{e-ph}}[\Delta,\vec{u}]=-\Tr\left(\Gamma_{\eta}\eta^2+\Gamma_{\phi}\phi^2\right)\delta\Delta,
\end{align}
where $\delta\Delta=(\kappa/\lambda)\Delta \text{div}\vec{u}$ is the change of the order parameter under the lattice strain and vertices
$\Gamma_{\eta,\phi}$ are defined by inverse propagators via  Ward-like identities:
\begin{equation}
\label{Gamma0}
	\Gamma_{\eta(\phi)}=\frac{\delta L_{\eta(\phi)}^{-1}}{\delta\Delta}.
\end{equation}
%

Using explicit forms of  the amplitude (\ref{L_eta}) and phase (\ref{L_phi}) propagators, one finds the part of action describing
electron-phonon interaction:
\begin{equation}
\begin{split}
	\label{A_e-ph}
	\A_{\text{e-ph}}[\eta,\phi,\vec{u}]=&\frac{\kappa_*}{W}\Tr\left(\frac{\partial(\Delta^2b)}{\partial\ln\Delta}\eta^2+\right.
	\\
	&\left.+\frac{v^2}{4}[(\nabla\eta)^2+(\nabla\phi)^2]\right)\text{div}\vec{u},
\end{split}
\end{equation}
where we had introduced an ``enhanced'' coupling constant $\kappa_*=\kappa/\lambda$ and $b\equiv b(i\Omega_n/2\Delta)$ 
is a function given by Eq.(\ref{a}).

\section{Ultrasonic attenuation}

\begin{figure}[t]
	\begin{minipage}[h]{0.49\linewidth}
		\center{\includegraphics[width=0.9\linewidth]{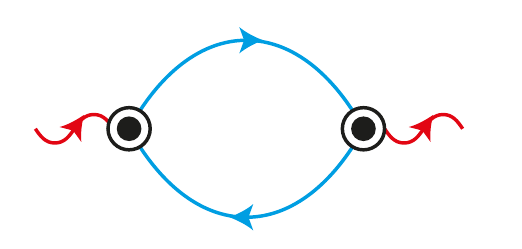}}
	\end{minipage}
	\hfill
	\begin{minipage}[h]{0.49\linewidth}
		\center{\includegraphics[width=0.9\linewidth]{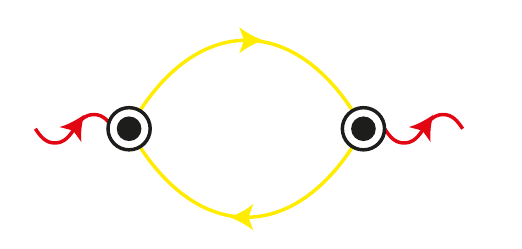}}
	\end{minipage}
	\caption{Ultrasonic attenuation due to collective superconducting modes. Left figure shows contribution of the amplitude mode, when a phonon (red line) falls apart into two amplitude modes (blue lines). Exactly same process happens via phase modes(yellow lines), shown on the right.}
\end{figure}

The ultrasonic attenuation $\alpha$ 
can be conveniently expressed via an acoustic $Q$-factor
\begin{equation}
	Q^{-1}=\frac{\alpha}{\omega}=\frac{1}{\rho_m\omega^2}\Im\Sigma^\text{A}_{\text{ph}},
\end{equation}
where $\Sigma^A_{\text{ph}}$ is an advanced phonon self-energy defined on  real frequencies.
Phonon spectrum is assumed to be acoustic, $\omega(q) = s q$,  with sound velocity $s \ll v$, i.e. the phonon propagator reads as
\begin{equation}
D^A(\omega,\vec{q})=\frac{1}{\rho_m[(\omega-i0)^2-s^2\vec{q}^2]-\Sigma^A_{\text{ph}}}.
\end{equation}

\subsection{Phase mode}

Ultrasonic attenuation due to interaction with the phase mode is given by processes in which acoustic phonon falls apart into two phasons,
\begin{equation}
\begin{split}
\Sigma_{\text{ph-$\phi$}}(i\omega_n,\vec{q})=\sumint_{m,\vec{k}}\abs{\Gamma_\phi(\vec{k},\vec{q})}^2\times
\\
\times L_{\phi}(i\Omega_m,\vec{k}) L_{\phi}(i\Omega_{m+n},\vec{k}+\vec{q}),
\end{split}
\end{equation}
where the electron phonon interaction vertex $\Gamma_\phi$ follows from Eq.(\ref{A_e-ph}),
\begin{equation}
	\Gamma_\phi(\vec{k},\vec{q})=iq\frac{\kappa_*\nu v^2}{4}[\vec{k}\cdot(\vec{k}+\vec{q})].
\end{equation}

\begin{equation}
\begin{split}
\Sigma_{\text{ph-$\phi$}}(i\omega_n,\vec{q})=\frac{\kappa_*^2\nu^2v^4}{16}q^2\sumint_{m,\vec{k}}[\vec{k}\cdot(\vec{k}+\vec{q})]^2\times
\\
\times L_{\phi}(i\Omega_m,\vec{k}) L_{\phi}(i\Omega_{m+n},\vec{k}+\vec{q}),
\end{split}
\end{equation}

Analytical continuation is performed in a standard fashion and gives for the imaginary part of phonon self-energy:
\begin{equation}
\begin{split}
\label{Sigma_phi}
\Im\Sigma_{\text{ph-$\phi$}}(\omega,\vec{q})=\frac{\kappa_*^2\nu^2v^4}{8}q^2\int_{\Omega,\vec{Q}}[\vec{k}\cdot(\vec{k}+\vec{q})]^2\times
\\
\times \left[B(\Omega+\omega)-B(\Omega)\right]\times
\\
\times\Im L_{\phi}^\text{R}(\Omega,\vec{k}) \Im L_{\phi}^\text{R}(\Omega+\omega,\vec{k}+\vec{q})
\end{split}
\end{equation}
with $B(x)=\coth x/2T$. Since 
\begin{equation}
\Im L_{\phi}^\text{R}(\Omega,\vec{k})=\frac{2\pi}{\nu vk}\Big(\delta(\Omega-vk)-\delta(\Omega+vk)\Big),
\end{equation}
as it follows from Eq.(\ref{L_phi}),
we eventually get 
\begin{equation}
Q^{-1}_{\text{ph}-\phi}=\frac{2\pi^4\kappa_*^2}{15}\frac{T^4}{\rho_msv^4},
\label{Qp1}
\end{equation}
where we had assumed that $\omega (v/s) \ll T$. 
 The result can also be rewritten as 
\begin{equation}
Q^{-1}_{\text{ph}-\phi}=\frac{2\pi^4}{15}\kappa^2\left(\frac{a_0}{R}\right)^4\left(\frac{T}{\Delta}\right)^4,
\label{Qp2}
\end{equation}
where $a_0=(\rho_ms)^{-1/4}$ by order of magnitude is equal to a material lattice constant and $R$ is a typical interaction radius.

\subsection{Amplitude mode}

The contribution of amplitude mode $\eta$ to the phonon decay rate is due to rare (at $T \ll \Delta$) collective excitation
with energies close to the threshold $\Omega = 2\Delta$.  In this energy region the electron-phonon interaction vertex is singular:
%
\begin{align}
	\Gamma_\eta(i\Omega_k)\simeq i\kappa_*\nu q\frac{\Delta^2 b\left(\frac{i\Omega_k}{2\Delta}\right)}{1+(\Omega_k/2\Delta)^2}.
\end{align}
where $\Omega$ stands for the energy of collective mode and phonon energy $\omega \to 0$.
This singularity  comes from the shift of the energy gap position $2\Delta$ induced by phonons via the lattice strain. 
For real frequencies close to the gap, we obtain for the absolute square of this vertex:
\begin{align}
	\abs{\Gamma_\eta(\Omega)}^2\simeq\frac{\pi^2}{4}\kappa_*^2\nu^2q^2\frac{\Delta^5}{\abs{\Omega-2\Delta}}.
\end{align}
If phonon energy $\omega$ becomes non-negligible, the singularity in the vertex $\Gamma$ is washed out as
\begin{align}
	\frac{1}{\Omega-2\Delta}\rightarrow\frac{1}{\sqrt{(\Omega-2\Delta)(\Omega+\omega-2\Delta)}}
\end{align}

The propagator of the amplitude mode is given by Eq.(\ref{L_eta_2});  in particular its imaginary part is equal to
\begin{equation}
\begin{split}
\Im L_{\eta}^\text{R}(\Omega,\vec{k})=\frac{4}{\nu} \frac{\gamma\Delta^2 }{ \Delta^4\gamma^2+v^4k^4},
\\
(\gamma=2\pi\sqrt{(\Omega-2\Delta)/\Delta})
\end{split}
\end{equation}
where it is assumed that  $0 < \Omega-2\Delta \ll\Delta$. 

Similarly to the phason contribution (\ref{Sigma_phi}), for the ultrasonic attenuation due to the acoustic phonon decay into two amplitude excitations we have
\begin{equation}
\begin{split}
\Im\Sigma_{\text{ph-$\eta$}}(\omega,\vec{q})=\int_{\Omega,\vec{Q}}\abs{\Gamma_\eta(\Omega,\omega)}^2\times
\\
\times \left[B(\Omega+\omega)-B(\Omega)\right]\times
\\
\times\Im L_{\eta}^\text{R}(\Omega,\vec{k}) \Im L_{\eta}^\text{R}(\Omega+\omega,\vec{k}+\vec{q}).
\end{split}
\end{equation}

%
This integral has an infrared divergence that is regularized by phonon frequency $\omega$, the leading contribution comes mainly from energies $(\Omega-2\Delta)\sim\omega$. Finally, for the amplitude contribution to the inverse quality factor we find
%
\begin{align}
\label{Qa}
\begin{split}
Q^{-1}_{\text{ph}-\eta}=&\frac{64\sqrt{\pi}}{3}\kappa_*^2\frac{\Delta^4}{\rho_ms^2v^3} \left(\frac{\Delta}{T}\right) \left(\frac{\omega}{\Delta}\right)^{3/4}e^{-2\Delta/T}
\end{split}
\end{align}
Comparing it with "phason" contribution (\ref{Qp1},\ref{Qp2}) we note that phase and amplitude mechanisms lead to
completely different behaviors in function of  ultrasound frequency $\omega$ and temperature $T$. In particular,
amplitude contribution scales as $\omega^{3/4}$ while the phase one is $\omega$-independent.
The ratio of both contributions is given by
\begin{equation}
\frac{Q^{-1}_{\text{ph}-\eta}}{Q^{-1}_{\text{ph}-\phi}}=\underbrace{\frac{480}{\pi^{7/2}}}_{=8.73} 
\left(\frac{v}{s}\right)\left(\frac{\Delta}{T}\right)^5 \left(\frac{\omega}{\Delta}\right)^{3/4}e^{-2\Delta/T}
\label{ratio}
\end{equation}
Asymptotically, at $T/\Delta \to 0$, phase contribution  dominates, although there is a range of relatively small $T/\Delta$,
 where the main contribution comes from the amplitude mode.
In any case, gapless phason mode is an artifact of the model with completely absent Coulomb interaction; as we will see in the next
 Section, even rather weak long-range Coulomb interaction pushes the energy of the lowest collective excitations to $2\Delta$.

The result (\ref{Qa}) can be compared to the similar formula from  Ref.~\cite{Att1},  with ultrasound attenuation in a 
usual s-wave superconductor leading to $Q^{-1} \propto exp(-\Delta/T)$.  The difference by factor 2 in the exponent is
important, and it is due to different statistical weights of excitations in two models: while in BCS theory independent
 electron and hole quasiparticles appear due to breaking of any Cooper pair,  a pseudospin superconductor supports 
 single-particle excitations with the lowest energy $2\Delta$.

\section{Role of Coulomb interaction}

In this Section we consider modifications caused by the long-range Coulomb interaction that was neglected previously 
in this paper.  The major effect of Coulomb interaction upon the low-temperature symmetry-broken state is formation of
the spectral gap $\Delta_\phi$ for the phase mode by the Anderson-Higgs mechanism.
We will see that $\Delta_\phi$  is usually large in comparison with the amplitude gap $\Delta$, thus the actual
threshold for all inelastic processes at $T=0$ is determined by $\Delta$.  For the same reason, at low $T$ 
total phonon decay rate is proportional to $\exp(-2\Delta/T)$, while power-law contribution (\ref{Qp2}) usually disappears.

We  consider effectively  3-dimensional problem and employ the  simplest way to introduce Coulomb interaction between electron pairs:
\begin{eqnarray}
\label{HC}
H[\vec S_i]=-2\sum_{i=1}^{N} (\xi_i+\Phi_i)S_i^z-\frac{1}{2}\sum_{i,j=1}^{N}J_{ij}\lb S_i^-S_j^++S_i^+S_j^-\rb+ \\
\nonumber
\sum_{i<j}\Phi_i\left(\frac{4\pi e^2}{\epsilon\abs{\vec{r}_i-\vec{r}_j}}\right)^{-1}\Phi_j,
\end{eqnarray}
where in the last term a matrix inversion is  implied.
Repeating the same steps used above to derive (\ref{A0}) from (\ref{H0}), we come now to the action of the following form:
\begin{widetext}
\begin{equation}
\A[\Delta]=-\Tr\lb \Delta^{*}J^{-1}\Delta\rb+
\Tr\ln\lb i\varepsilon_l+(\xi+\Phi)\sigma^z +\frac{1}{2}(\Delta \sigma^-+\Delta^* \sigma^+) \rb
-\Tr\lb \Phi_i\left(\frac{4\pi e^2}{\epsilon\abs{\vec{r}_i-\vec{r}_j}}\right)^{-1}\Phi_j\rb.
\end{equation}
  In the approximation of constant DoS,
  phase and amplitude modes are decoupled, and the whole effect of Coulomb interaction is to
 provide a mass to a previously gapless phase mode due to mixing between $\Phi$ and $d\phi/dt$.  
The corresponding part of the action reads
\begin{equation}
\label{A_Coulomb}
\A[\phi,\Phi]=-\nu\Delta^2\Tr\lb \left(-\frac{v^2k^2}{4\Delta^2} + \Pi_{yy}\right)\phi^2 +2\Pi_{yz}\phi\cdot\left(\frac{\Phi}{\Delta}\right)
+\left(\Pi_{zz}+\nu^{-1}V^{-1}(k)\right)\cdot\left(\frac{\Phi}{\Delta}\right)^2
\rb
\end{equation}
\end{widetext}
where $V(q)=4\pi e^2/\epsilon q^2$ is Coulomb propagator and semionic polarization functions
$\Pi_{\alpha\beta}$, with $\alpha,\beta \in (y,z)$ are defined similar to Eq.(\ref{Pix});
calculation of the trace over semionic modes leads to
\begin{align}
	\label{Pi_yy}
\Pi_{yy}(x)=-\frac{x}{\sqrt{1+x^2}}\text{arcsh}\,x,
\\
\label{Pi_yz}
\Pi_{yz}(x)=i\frac{1}{\sqrt{1+x^2}}\text{arcsh}\,x,
\\
\label{Pi_zz}
\Pi_{zz}(x)=\frac{1}{x\sqrt{1+x^2}}\text{arcsh}\,x,
\end{align}
where $x=\Omega_n/2\Delta$.
Next we integrate out electric potential $\Phi$ and obtain the phase-only action in the form 
\begin{equation}
\label{AC2}
\A_\text{eff}[\phi]=
-\nu\Delta^2\int_{\xi}\Tr\lb \phi\left(-\frac{v^2k^2}{4\Delta^2} + \Pi_{yy}-\frac{\Pi_{yz}^2}{\Pi_{zz}+\nu^{-1}V^{-1}}\right)\phi
\rb.
\end{equation}
Now we substitute expressions for $\Pi_{\alpha\beta}$ into Eq.(\ref{AC2}) and obtain inverse phase  propagator in the form
\begin{widetext}
\begin{equation}
\label{LC1}
L_\phi^{-1}(i\Omega_n)=\frac{\nu v^2k^2}{4}\left[1+\frac{\epsilon W\Omega_n^2}{4\pi e^2v^2}\left(1+\left(\frac{\epsilon W k^2}{4\pi e^2}\right)\frac{\frac{\Omega_n}{2\Delta}\sqrt{1+\left(\frac{\Omega_n}{2\Delta}\right)^2}}{\text{arcsh}\,\frac{\Omega_n}{2\Delta}}\right)^{-1}\right].
\end{equation}
This expression should be analytically continued to the real energy axis, $i\Omega_n\rightarrow\Omega+i0$, which leads to
\begin{equation}
\label{LC2}
\left(L_\phi^{R}(\Omega,k)\right)^{-1}=\frac{\nu v^2k^2}{4}\times
\begin{cases}
	1-\frac{\epsilon W\Omega^2}{4\pi e^2v^2}\left(1+\left(\frac{\epsilon W k^2}{4\pi e^2}\right)\frac{\frac{\Omega}{2\Delta}\sqrt{1-\left(\frac{\Omega}{2\Delta}\right)^2}}{\text{arcsin}\,\frac{\Omega}{2\Delta}}\right)^{-1} & \Omega<2\Delta
	\\
	1-\frac{\epsilon W\Omega^2}{4\pi e^2v^2}\left(1+\left(\frac{\epsilon W k^2}{4\pi e^2}\right)\frac{\frac{\Omega}{2\Delta}\sqrt{\left(\frac{\Omega}{2\Delta}\right)^2-1}}{\ln\left[\frac{\Omega}{2\Delta}+\sqrt{\left(\frac{\Omega}{2\Delta}\right)^2-1}\right]-i\frac{\pi}{2}}\right)^{-1} & \Omega\geq2\Delta
\end{cases}.
\end{equation}
\end{widetext}
Propagator of the phase mode (\ref{LC2}) possesses (at $k=0$)  two types of singularities in the complex plane
 of $\Omega$:  a branch cut  along $(-\infty,-2\Delta]\cup[2\Delta,+\infty)$ and  a simple pole  at the plasmon frequency
\begin{equation}
\Delta_\phi = \sqrt{4\pi e^2\nu v^2/\epsilon }
\label{DeltaPhi}
\end{equation}
At small momenta $k$ the pole (\ref{DeltaPhi})  shifts according to $\Omega(k) = \sqrt{\Delta^2_\phi  +v_{\phi*}^2k^2}$
and 
$v_{\phi*}=v\times\frac{\Delta_\phi}{2\Delta}\left(\ln\frac{\Delta_\phi}{\Delta}\right)^{-1/2}$.

The key point for further analysis is the relation between amplitude gap $\Delta$ and plasmon gap $\Delta_\phi$.
For the phase mode to be well-defined at $k >0$, its energy gap $\Delta_\phi$ should be below $2\Delta$, otherwise
imaginary part appears in the inverse propagator (\ref{LC2}) at the mass shell of the massive phase mode.
In such a case, actual threshold for all inelastic processes is given by $2\Delta$ and phase mode itself is irrelevant.

Estimates below show that usually $\Delta_\phi  \geq 2\Delta$ indeed. Instead of using
model-dependent relation (\ref{DeltaPhi}), we rewrite $\Delta_\phi$  in terms of observables:
\begin{equation}
\Delta_\phi^2 = \frac{4\pi e^2}{\epsilon} \frac{\hbar^2 \rho_s}{e^2}
\label{DeltaPhi2}
\end{equation}
where $\rho_s$ is the superfluid density defined via  London relation for supercurrent,
 $\vec{j} = -\rho_s \vec{A}/c$.  Next, we use the estimate~\cite{FI2015} for superfluid density
of pseudogapped superconductor, $\rho_s \sim \nu_0 e^2 R^2 \Delta^2/\hbar^2 $, to obtain
\begin{equation}
\frac{\Delta_\phi^2}{4\Delta^2} \approx \frac{\pi e^2}{\epsilon} \nu_0 R^2
\label{ratio}
\end{equation}
Although the model used in~\cite{FI2015} is different from our present one (here we employ a
 very large interaction range $R$ to use mean-field approximation, while in~\cite{FI2015} hopping of 
pairs via Mott-type pair resonances was  assumed), the results for the ratio $\Delta_\phi/\Delta$
are similar in both models.

To get some feeling of relevant numbers, we use parameters known for amorphous superconducting InO$_x$:
for the density of states we take~\cite{Zvi} an estimate $\nu_0 \sim 2 \cdot 10^{33} erg^{-1}cm^{-3} $;
 for the estimate of effective hopping range $R$ we can use the value of superconducting coherence length
$\xi_0 \approx 4-5 $ nm extracted from $H_{c2}$ measurements in less disordered superconducting InO$_x$
in Ref.~\cite{Sacepe2015}.  Combining all together, we find $\approx 500/\epsilon$ for the R.H.S. of
Eq.(\ref{ratio}). Unfortunately,  effective dielectric constant of InO$_x$ in the insulating phase  was
not yet measured. 

Another approach to the problem one can try is a purely theoretical one:  consider
non-interacting Anderson insulator with a localization length $L_{loc}$ and find its dielectric response
at $T=0$.  Such a program was realized recently~\cite{FIC2016} numerically; for the 3D case it results
in $\epsilon \approx 3 e^2 \nu_0 L^2_{loc}$.  Substituting this estimate into Eq.(\ref{ratio}), one 
finds surprisingly simple and universal result:  $\Delta_\phi/2\Delta \approx R/L_{loc} > 1$.
Thus we conclude that most probably plasmon gap is too large for the phase mode to be relevant for 
ultrasound decay.

However, the above conclusion can be incorrect for some special highly polarizable materials 
with a very high intrinsic dielectric constant, like SrTiO$_3$ with its $\epsilon > 10^4$.
Very light doping of  SrTiO$_3$  makes it superconducting~\cite{SrTiO3,SrTiO3new}.
Such a superconductor may  have unusually small phason gap $\Delta_\phi \ll \Delta$; in this case
the results (\ref{Qp1},\ref{Qp2}) for ultrasound decay into phase mode might be relevant.


\section{Conclusions}

We have shown in this paper that  phonon decay rate in pseudogapped superconductor at low temperatures
$T \ll T_c$ is determined by its collective modes. 
The amplitude mode  has a threshold energy $2\Delta < \Delta_P$, so its contribution to the
decay of low-frequency phonons is given by  Eq.(\ref{Qa}). 
It is proportional to $\exp(-2\Delta/T)$ and dominates over "usual" \,  single-particle contribution which 
is $\propto \exp(-\Delta_P/T)$.  Thus measurements of ultrasound attenuation rate may provide an additional
way to determine the value of the collective gap.

If Coulomb interaction between conduction electrons
 is very strongly suppressed due to high intrinsic
dielectric constant $\epsilon \geq 10^3$, an additional contribution from the phase 
mode may be present, see Eq.(\ref{Qp2}). 
 Different dependences of the amplitude and phase contributions to the decay rate on frequency and temperature make 
it possible to  identify both contributions  separately.

We are grateful to L. B. Ioffe and V. E. Kravtsov for useful discussions. 
  Research of M.V.F. was partially  supported by the Russian Science Foundation grant
\#  14-42-00044. The research was also
partially supported by the RF Presidential Grant No.
NSh-10129.2016.2.

\end{document}